\pdfoutput=1
\documentclass[twocolumn,prb,aps,showpacs,numbers,amsmath,amssymb,superscriptaddress,floatfix]{revtex4-1}
\usepackage{graphicx}
\usepackage{dcolumn}
\usepackage{bm}
\usepackage{epsf}
\usepackage{hyperref}

\begin{document}

\title{Hypergeometric resummation of self-consistent sunset diagrams for electron-boson quantum many-body systems out of equilibrium}
\author{H\'ector Mera}
\affiliation{Department of Physics and Astronomy, University of Delaware, Newark, DE 19716-2570, USA}
\author{Thomas~G. Pedersen} 
\affiliation{Department of Physics and Nanotechnology, Aalborg University, DK-9220 Aalborg East, Denmark}
\author{Branislav K. Nikoli\'{c}}
\affiliation{Department of Physics and Astronomy, University of Delaware, Newark, DE 19716-2570, USA}

\begin{abstract}
A newly developed hypergeometric resummation technique [H. Mera {\it et al.}, Phys. Rev. Lett. {\bf 115}, 143001 (2015)] provides an easy-to-use recipe to obtain conserving approximations within the self-consistent nonequilibrium many-body perturbation theory. We demonstrate the usefulness of this technique by calculating the phonon-limited electronic current in a model of a single-molecule junction within the self-consistent Born approximation for the electron-phonon interacting system, where the perturbation expansion for the nonequilibrium Green function in powers of the free bosonic propagator typically consists of a series of non-crossing ``sunset'' diagrams. Hypergeometric resummation preserves conservation laws and it is shown to provide substantial convergence acceleration relative to more standard approaches to self-consistency. This result strongly suggests that the convergence of the self-consistent ``sunset'' series is limited by a branch-cut singularity, which is accurately described by Gauss hypergeometric functions. Our results showcase an alternative  approach to conservation laws and self-consistency where expectation values obtained from conserving perturbation expansions are ``summed'' to their self-consistent value by analytic continuation functions able to mimic the convergence-limiting singularity structure. 
\end{abstract}

\maketitle

\section{Introduction}
In contrast to bulk materials, nanostructures can be easily brought into far from equilibrium states by applying a bias voltage or external time-dependent fields.~\cite{Kamenev2011} Simulation  tools~\cite{Haug2008,Stefanucci2013,Brandbyge2002,Taylor2001,Palacios2002,Areshkin2010,Pecchia2004,Lake1997,Kubis2011}  make it possible to evade usual trial-and-error experimental procedures by screening nanostructures {\it in silico} with desired properties for applications. The traditional semiclassical simulation tools  cannot be used for quasiballistic nanometer-size active region attached to much larger reservoirs. On the other hand, accounting for all the relevant quantum many-body interactions for such systems composed of thousands of atoms is computationally prohibitively expensive.~\cite{Rhyner2014,Luisier2014}  For example, simulation of  photocurrent in a photovoltaic cell requires to take into account electronic structure of the cell, electron-photon interactions responsible for photoexcitation, by electron-hole recombination processes, emission and absorption of phonons and scattering by disorder.~\cite{Aeberhard2012} Spintronic devices, such as spin-transfer torque magnetic random access memory,~\cite{Locatelli2014} furnish another example of a complex quantum many-body system where electrons, magnons and phonons interact with each other while being driven away from equilibrium.~\cite{Manchon2009a,Levy2006a,Mahfouzi2014} 

The nonequilibrium Green's function formalism (NEGF)~\cite{Haug2008,Stefanucci2013} provides a rigorous framework to model such systems by  extending the many-body perturbation theory (MBPT) to out of equilibrium regimes. In this approach, any one-particle observable  is calculated from the one-particle NEGF by solving the Dyson equation. In practice, unfortunately, the resulting equations need to be approximated. Typical approximations involve the use of finite-order perturbation theory,~\cite{Viljas2005,Paulsson2005,Frederiksen2007,Mera2012,Mera2013,Cavassilas2013,Bescond2013} where  the contour-ordered NEGF is approximated by a finite number of diagrams; or partial resummation schemes, like the $GW$ approximation,~\cite{Thygesen2008,Spataru2009,Dash2010,Dash2011,Tandetzky2015} where one class of diagrams (or a few of them) is ``summed to infinite order.''  Since there are infinitely many such classes, each containing an infinite number of diagrams, both finite-order PT and ``infinite-order'' resummation schemes are tentative at best.  Furthermore the NEGF equations and their approximations are nonlinear integral equations which require a self-consistent solution,~\cite{Mahfouzi2014,Mera2012} thus making their numerical solution very demanding from a computational perspective.~\cite{Frederiksen2007,Luisier2014,Mera2013} 

Nevertheless both finite-order PT and infinite-order partial resummations are widely used to simulate out of equilibrium systems in the presence of interactions. The so-called self-consistent Born approximation (SCBA)~\cite{Stefanucci2013,Lee2009a} is a Hartree-Fock-like approximation commonly employed to model electron-electron, electron-phonon, electron-photon and electron-magnon interactions. The SCBA is the simplest self-consistent non-crossing approximation that is also $\Phi$-derivable and, therefore, conserving.~\cite{Stefanucci2013} Despite its simplicity,  atomistic simulations using the SCBA can be very challenging, particularly when realistic system sizes are considered and a thorough exploration of the space of device parameters  is intended.~\cite{Rhyner2014,Luisier2014,Frederiksen2007} 

\begin{figure}
\includegraphics[scale=0.35]{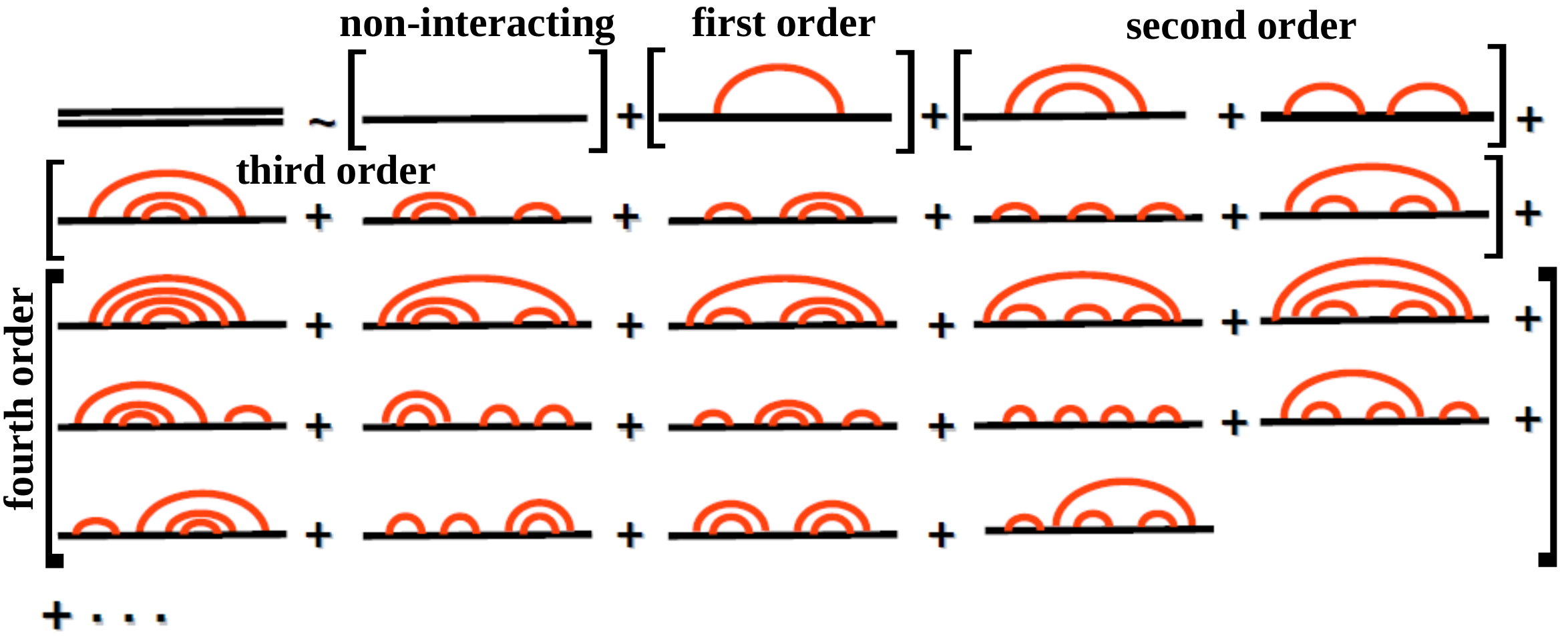}
\caption{(Color online) Diagrammatic representation of the interacting NEGF (double straight line) within the Fock-SCBA. This NEGF is asymptotic to a series in  powers of the noninteracting NEGF (single straight line) and the boson propagator (semicircle). The brackets enclose all ``sunset'' diagrams contributing to each order in the expansion in powers of  $U^2$  (square of the strength of electron-boson interaction).}
\label{fig:fig0} 
\end{figure}

The purpose of this study is to demonstrate an alternative theoretical and computational approach where excellent approximations to the self-consistent results are obtained by combining finite-order PT with a recently-developed hypergeometric resummation scheme~\cite{Mera2015,Pedersen2015}. For the sake of simplicity, we will use the Fock only SCBA (Fock-SCBA) as a prototype of self-consistent partial resummation approximation which, as illustrated in Fig.~\ref{fig:fig0}, consists of a series of self-consistent ``sunset'' diagrams. Nevertheless, we expect our insights to be useful for the calculation of quantities under other more involved (but equally uncontrolled) self-consistent resummation schemes.~\cite{Thygesen2008,Spataru2009,Dash2010,Dash2011}  While PT {\it per se} is, at best, an intrinsically weakly-interacting approach, its combination with a carefully crafted analytic continuation  function (ACF) can yield accurate results far beyond the weakly interacting limit, even allowing the calculation of intrinsically non-perturbative quantities from low-order  PT.~\cite{Mera2015,Pedersen2015,Sanders2015}  

For this purpose, hypergeometric functions $_pF_q$ are good candidates for ACF since they can mimic various types of singularities responsible for the divergence of a perturbation expansion. Thus far, hypergeometric resummation has only been applied to a handful of problems---originally it was tested using $_2F_1$ on the examples of divergent perturbation series in single particle quantum mechanics.~\cite{Mera2015,Pedersen2015} The same approach was subsequently utilized in Ref.~\onlinecite{Sanders2015} for the calculation of the critical exponents in the two-dimensional Bose-Hubbard model where higher-order hypergeometric functions $_pF_q$ were also considered.  In these cases hypergeometric resummation revealed itself as a method superior to widely used resummation approaches, such as Shanks transformation,~\cite{Caliceti2007} Pad\'e approximants~\cite{Baker1996} and Borel-Pad\'{e} resummation.~\cite{Caliceti2007}

Here we demonstrate that hypergeometric resummation can be used as a convergence acceleration scheme for the calculation of expectation values from the self-consistent solution of the Dyson equation in nonequilibrium MBPT. It will be argued by example that, relative to standard   solvers for the Dyson equation in nonequilibrium MBPT, hypergeometric resummation has great potential to drastically reduce the computational cost of calculations without significantly impacting their accuracy. The rest of the paper is organized as follows. In Sec.~\ref{sec:dyson}, we overview  the standard approach to solve the Dyson equation self-consistently. Section~\ref{sec:pt} discusses perturbative approximations to the NEGF and Pad\'e approximants as ACFs. In Sec.~\ref{sec:hyper}, we review the hypergeometric approximants from Ref.~\onlinecite{Mera2015} and we  introduce a new flavor of hypergeometric resummation based on the ratio test for series convergence. Section~\ref{sec:junction} introduces a physically motivated model of a single-molecule junction to which we apply various techniques discussed in previous sections by calculating of the phonon-limited electronic current. A comparison with both Pad\'e approximants and standard iterators reveals that hypergeometric resummation outperforms both of these approaches. In Sec.~\ref{sec:discussion} we discuss these results and argue that the self-consistent ``sunset'' series has a generally finite radius of convergence and that its divergence is due to a branch-cut in an abstract plane of complex values for the electron-phonon interaction strength parameter. We conclude in Sec.~\ref{sec:conclusion}.

\section{The self-consistent Dyson equation}\label{sec:dyson}

The central objects~\cite{Stefanucci2013} of NEGF formalism are the one-particle contour-ordered GF, $G$, and self-energy $\Sigma=\Sigma[G]$. The latter is a functional of $G$ that takes into account the effect of interactions and, in practice, needs to be approximated.  $G$ and $\Sigma[G]$ are related by the self-consistent Dyson equation
\begin{equation}\label{eq:dyson}
G=g+g\Sigma[G]G\,,
\end{equation}
which is a short-hand notation for 
\begin{equation}
G(1,2)=g(1,2)+\int\,d3 d4\,g(1,3)\Sigma(3,4)G(4,2).
\end{equation}
Here $g$ is the noninteracting contour-ordered NEGF and $i=(\sigma_i,\mathbf{r}_i,t_i)$ is a global index encompassing spin, position and time. The time arguments are located on the Keldysh-Schwinger contour, consisting of two counter-propagating copies of the real-time axis (the forward branch extending from $-\infty$ to $\infty$ and the backward branch extending from $\infty$ to $-\infty$), which is the hallmark of the NEGF formalism.  Using the short-hand notation, we can also rewrite Eq.~\eqref{eq:dyson} as
\begin{equation}
G=[g^{-1}-\Sigma[G]]^{-1}\,.
\end{equation}
The Dyson equation must be solved for $G$ and is explicitly self-consistent, thereby calling for an iterative scheme. The standard iteration  proceeds as follows: one approximates $G\approx g$ and calculates $\Sigma[g]$ to build a new approximation $G_1=[g^{-1}-\Sigma[g]]^{-1}$, then one approximates $G\approx G_1$ and evaluates $\Sigma[G_1]$. This procedure is then repeated typically according to the expression
\begin{equation}\label{eq:iterative}
G_n=[g^{-1}-\Sigma[G_{n-1}]]^{-1}\,,
\end{equation}
until $G_n$ and $G_{n-1}$ (or the relevant expectation values calculated from them) differ by less than some prescribed tolerance. This iterative procedure does not guarantee convergence and often requires convergence acceleration schemes, such as linear and Pulay mixing which typically need to be carefully combined with a preconditioning scheme like Nieminen-Kerker~\cite{Nieminen1977,Kerker1981} or direct inversion of the iterative subspace~\cite{Kresse1996} without, again, guaranteeing convergence.  

The appeal of self-consistent approaches over simpler forms of PT (to be discussed below) is based on the relationship between  conservation laws and self-consistency~\cite{Stefanucci2013,Mera2013,Baym2000}---it is well known that by choosing a conserving self-energy approximation, the expectation values calculated from the self-consistent $G$ will obey whatever conservation laws they ought to obey, i.e. the fully self-consistent $G$ is a conserving GF. However, as pointed out by Baym,~\cite{Baym2000,Baym1962} full self-consistency is by no means a necessary condition for conservation laws to be obeyed---{\it a perturbation expansion in powers of the interaction strength will also be conserving at each and every order, provided that one chooses a conserving self-energy and keeps all the resulting terms at each order}.~\cite{Mera2012,Mera2013} In contrast, an approximation like $G_1$ obtained by setting $n=1$ in Eq.~\eqref{eq:iterative} above is therefore not conserving because it misses some of the second-order diagrams owing to the fact that the perturbative expansion for the NEGF is not a geometric series [as incorrectly assumed in Eq.~\eqref{eq:iterative}]. While conserving, perturbative approximations are valid only for weakly interacting systems and break down very rapidly as the interaction gets stronger than the radius of convergence of the perturbative  expansion~\cite{Caliceti2007}. Thus, being ``conserving'' is not a good enough reason to choose an approximation for $\Sigma$. On the other hand, as we will demonstrate, being ``perturbative'' is not a good enough reason to discard an approximation.

 In fact, critically reflecting on the basis of self-consistent approximations in MBPT is likely to raise some tough questions: it appears that the  limit of validity of this method is, strictly speaking, identical to that of the underlying perturbation series. The reason for this is the neglect of---or the need to approximate---vertex corrections,~\cite{Dash2010,Dash2011,Tandetzky2015} which are present already at the second order in PT in the form of a first order vertex. If one measures the accuracy of an approximation by the fraction of Feynman diagrams it accounts for, one readily sees that at high orders essentially all the Feynman diagrams include high-order vertex diagrams.~\cite{Molinari2005,Molinari2006} Any form of self-consistent PT is an infinite-order approximation where, by including a finite-order approximation to the vertex, one is summing an error to infinite order.~\cite{Gukelberger2015} But $\Sigma$ is supposed to correct $g$---if the error is  not small compared to the correction then self-consistent MBPT is not applicable; if it is small then a finite-order, perturbative approach is likely to be equally applicable. Accordingly, we see no fundamental reason for self-consistent MBPT to be favored over simpler forms of PT---if it works its perturbation series will, most likely, also work.~\cite{Mera2012,Mera2013} 
 
 Nevertheless, the use of self-consistent MBPT may be justified to some extent. When a convergent self-consistent approximation is obtained it is typically \emph{regularized}, i.e., it is free from the absurdly large values that one obtains from finite-order PT outside its radius of convergence. There are also more mundane reasons---we would like to have as many approximations as possible in our toolbox. Analytic continuation and resummation techniques may be very helpful in this respect because, as shown below, they may provide substantial convergence acceleration relative to state-of-the-art self-consistent solvers, which are based on the combination of Eq.~\eqref{eq:iterative} with mixers and preconditioner.  

\section{Perturbative NEGF and Pad\'e Resummation}\label{sec:pt}

An alternative to Eq.~\eqref{eq:iterative} is to compute the perturbation series for the NEGF, which is conserving if $\Sigma$ is a conserving approximation, and compute expectation values from it. This yields a perturbation expansion for the expectation value. The perturbation expansion for the expectation value can then be analytically continued as analytic continuation naturally preserves the conservation laws. 

Let us begin by considering (say) an interacting many electron system, where the interaction is mediated by the exchange of bosons and described by the usual Feynman diagrams. To each vertex at the edges of a bosonic propagator we ascribe a parameter $U$ which controls the interaction strength. The perturbation expansion of $G$ in powers of $U^2$ up to order $N$ is then given by
\begin{equation}\label{eq:gn}
g_N=g+\sum_{n=1}^N \delta g_n U^{2n},
\end{equation}
where $\delta g_i$ are the expansion coefficients that do not depend of $U$. Therefore, perturbative NEGFs are polynomials in $U^2$, as illustrated for Fock-SCBA in Fig.~\ref{fig:fig0}.

A simple expression for $g_N$ in terms of $\Sigma$ is
\begin{equation}\label{eq:gnpt}
g_N=g_{N-1}+g\sum_{n=1}^N \Sigma_n\Delta g_{N-n},
\end{equation}
where $\Delta g_n=\delta g_n U^{2n}$, $\delta g_0 = g_0 =g$ and $\Sigma_n$ contains all the (proper) self-energy diagrams with $n$ bosonic  lines. Furthermore if, like in the case of the  SCBA, $\Sigma$ is a linear functional of $G$ then $\Sigma_n=\Sigma[\Delta g_{n-1}]$:  in this case it is straightforward to iterate Eq.~\eqref{eq:gnpt} a couple of times. For $n=1$ we get $g_1=g+g\Sigma_1 g=g+\Delta g_1$; for $n=2$ we get $g_2=g_1+g\Sigma_1\Delta g_1+g\Sigma_2 g$, and so on. Clearly we are getting the full perturbative expansion---terms coming from self-consistency (like $g\Sigma_1\Delta g_1$) and terms coming from $\Phi$-derivability (like $g\Sigma_2 g$). Therefore, as discussed in Ref.~\onlinecite{Mera2013}, $g_N$ is actually conserving, albeit not fully self-consistent. Indeed, $g_n$ contains a finite number of diagrams, while the fully self-consistent $G$ contains an infinite number of them. The diagrammatic representation of $g_4$ in Fock-SCBA is shown in Fig.~\ref{fig:fig0}.

As pointed out above, one could now take the sequence of $g_n$ apply to it some sequence transformation such as Pad\'e or Shanks. Unfortunately, as $G_1$ demonstrates, that does not typically lead to conservation laws. Instead these transformations can indeed be used as analytic continuation formulas, but on the sequence of expectation values obtained from the sequence of $g_n$. We can view $G_1$ as the 0/1 Pad\'e  approximant to the NEGF whose perturbation expansion is in general an unphysical geometric series that breaks conservation laws---it misses, for instance, the  $g\Sigma_2 g$ discussed above.

In NEGF theory, any one-electron expectation value, $\mathcal{O}$, is a functional of the NEGF, $\mathcal{O}=\mathcal{O}[G]$. One can then evaluate the expectation value of an observable using $g_n$ obtained from Eq.~\eqref{eq:gnpt}, $\mathcal{O}[g_N]$, thus generating a perturbation series for the expectation value to order $N$
\begin{equation}
 \mathcal{O}_N\equiv \mathcal{O}[g_N]=\sum_{n=0}^N \delta o_n U^{2n}=\mathcal{O}_0+\sum_{n=1}^N\Delta\mathcal{O}_n.
\end{equation} 
Here $\delta o_n=\mathcal{O}[\delta g_n]$ are the expansion coefficients, $\Delta\mathcal{O}_n=\delta o_n U^n$ is the $n$-th order correction to the noninteracting expectation value, and we are assuming that the expectation value is a linear functional of $G$. The Born series has a radius of convergence $U_c$. For $U<U_c$ one finds $\lim_{N\rightarrow \infty} \mathcal{O}_N \rightarrow \mathcal{O}[G]$, while for $U>U_c$  the sequence  $\mathcal{O}_N$ never converges to $\mathcal{O}[G]$ as $N$ increases. In fact, increasing $N$ for $U>U_c$ will only make things 
worse---the calculated expectation value either grows without bound or oscillates wildly between large negative and positive numbers.~\cite{Caliceti2007}  

However,  sequence transformations~\cite{Weniger1989,*Weniger2010} like the Pad\'e technique can actually accelerate convergence.  For instance, given $\mathcal{O}_1=\mathcal{O}_0+\Delta\mathcal{O}_1$ and $\mathcal{O}_2=\mathcal{O}_1+\Delta\mathcal{O}_2$, we can build the $1/1$ Pad\'e approximant
\begin{equation}
\mathcal{O}_{1/1}=\frac{\mathcal{O}_0+(\Delta\mathcal{O}_1^2-\mathcal{O}_0\Delta\mathcal{O}_2)/\Delta\mathcal{O}_1}{1-\Delta\mathcal{O}_2/\Delta\mathcal{O}_1},
\end{equation}
as well as the 0/2  Pad\'e approximant. If instead we have the sequence of expectation values up to $N=4$,  we can build, e.g., $2/2$ Pad\'e approximant, but the equation is too long to be written here.

It has been shown~\cite{Mera2013} that by buiding a Pad\'e table of expectation values one can achieve substantial convergence acceleration relative to the standard technique of calculating the sequence of $\mathcal{O}[G_n]$. However, there are some clear limitations---Pad\'e resummation approximates the exact expectation value by a rational functions of $U$. Furthermore, the denominator of a given Pad\'e approximant may vanish for specific values of the device parameters, rendering the approximation unusable and requiring the computation of higher orders in the perturbation expansion. 

Ultimately, the root cause of the divergence of a perturbation expansion is a singularity in an abstract plane~\cite{Dyson1952,Vainshtein2002,Bender1973,Suslov2005} of complex values of $U$ (the physical values are located on the real axis). Pad\'e approximants are able to describe the case where the convergence-limiting singularities are poles. But the singularity structure could also be a branch cut (a dense line of poles), and in such cases Pad\'e approximants lack the necessary analytic structure to describe the physical $U$-dependence, thus converging very slowly with perturbation order. In problems where convergence is limited by a branch cut, Pad\'e approximants quite literally attempt to reconstruct the cut by putting poles next to each other.~\cite{Caliceti2007} Accordingly the Pad\'e sequence should converge slowly, if at all, for those cases. Simple low-order approximants able to account for both branch-cut and poles should be advantageous.~\cite{Mera2015}

\section{Hypergeometric Resummation}\label{sec:hyper}

The combination of MBPT and analytic continuation of the expectation values calculated from it is a very promising and not widely explored approach to the computation of nonequilibrium properties of quantum many-body systems. It is naturally conserving and may well be a better option than the more standard methods, based on Eq.~\eqref{eq:iterative}. Unfortunately, in typical sequence transformations such as Shanks or Pad\'e, one approximates the $U$-dependence of the expectation value by a rational function of $U$. Therefore it is highly advisable to develop resummation techniques able to deal with both rational and non-rational functions of $U$, as well as with both poles and branch cuts in the complex  $U$-plane.

Here we put forward hypergeometric functions, in particular Gauss $_2F_1$ hypergeometric function, as tools for analytic continuation of expectation values calculated from MBPT. There are various reasons for this particular choice of analytic continuation functions: they are flexible and very general, including many other functions (such as binomials, exponentials, square roots, Bessel functions, ...) as particular cases; they are able mimic  both cuts and poles and, thus, can naturally model the two main types of convergence-limiting singularity structures; their Taylor expansions are known, allowing one to build hypergeometric approximants in a fashion akin to Pad\'e approximants.

Here we discuss two possible approaches to hypergeometric resummation. We first describe the hypergeometric approximant put forth in Ref.~\onlinecite{Mera2015}, and then introduce an alternative approach based on the ratio test of series convergence. In the first case, we seek approximations of the form
\begin{equation}\label{eq:hyper1}
\mathcal{O}= \,_2F_1(h_1,h_2,h_3,h_4 U^2)\,\mathcal{O}_0,
\end{equation}
while in the second case, one is lead to approximations of the form
\begin{equation}
\mathcal{O}=\mathcal{O}_0+ \, _2F_1(1, h_2^\prime,  h_3^\prime,  h_4^\prime U^2)\,\Delta \mathcal{O}_1.
\end{equation}
Here $h_i$ and $h_i^\prime$ parameters are to be determined from the perturbation coefficients for $\mathcal{O}$, in a way that is fashioned after Pad\'e resummation, by equating order-by-order the Taylor series for the hypergeometric approximants with the perturbation expansion for the observable. Fixing the parameters of both of these hypergeometric approximants requires four orders of PT. Thus, in the former approach one needs to determine the coefficients $h_{1-4}$, while in the latter one needs the first order correction $\Delta \mathcal{O}_1$, as well as the next three corrections, $\Delta \mathcal{O}_{2-4}$, in order to determine $h^\prime_{2-4}$. 

\subsection{Hypergeometric Approximant}\label{sec:hyperapprox}

Let us commence by recalling the construction of the hypergeometric approximant in Ref.~\onlinecite{Mera2015}.  First, the perturbation series for the expectation value of the observable under consideration is normalized by dividing it by its unperturbed value
\begin{equation}
\mathcal{O}/\mathcal{O}_0=1+o_1U^2+o_2U^4+\cdots,
\end{equation}
where $o_n=\delta o_n/\mathcal{O}_0$. The Taylor series for $_2 F_1$ is
\begin{equation}
_2 F_1(h_1,h_2,h_3;h_4 U^2)= \sum_{n=0}^\infty \frac{(h_1)_n(h_2)_n}{n! (h_3)_n} h_4^n U^{2n},
\end{equation}
where $(h_i)_n=\Gamma(h_i+n)/\Gamma(h_i)$ is a so-called Pochhammer symbol, defined in terms of Euler Gamma function. To obtain the $h_i$ that determine the hypergeometric approximant, one equates each order in the asymptotic series for $_2F_1$  with the corresponding term in the perturbation expansion for $\mathcal{O}$, resulting in a set of four (non-linear) equations with four unknowns $h_{1-4}$
\begin{equation}
o_n=\frac{(h_1)_n(h_2)_n}{n! (h_3)_n} h_4^n,\, 0<n\le 4\,. 
\end{equation}
Because the equations are non-linear, multiple solutions are possible. In the numerical example given below, however, two solutions were found, corresponding to the same hypergeometric function (see also Ref.~\onlinecite{Mera2015}). Once the $h_i$ have been obtained one has a hypergeometric approximant of the form given by Eq.~\eqref{eq:hypergeometric1}. The hypergeometric function plays the role of a $U$-dependent multiplicative factor which modulates the non-interacting expectation value bringing it, hopefully, in close agreement with the interacting result.

\subsection{Hypergeometric Approximants from the ratio test}

Another flavor of hypergeometric resummation is inspired by the ratio test of series convergence. A useful way to know in many instances whether a series converges or not is the ratio test---one just needs to look at the ratio between consecutive expansion coefficients in the perturbation series for the expectation value, $r(n)=\delta o_n/ \delta o_{n-1}$. If the ratio goes to a finite constant as $n \rightarrow \infty$ then the radius of convergence is not zero.

Let us then assume the radius of convergence to be not zero and approximate the ratio $r(n)$ between consecutive coefficients by a rational function---a diagonal Pad\'e approximant
\begin{equation}
r(n)\equiv \frac{\delta o_n}{\delta o_{n-1}} \approx \frac{p_0+p_1 n+\cdots+p_N n^N}{1+q_1n+\cdots q_N n^N}.
\label{eq:HAC}
\end{equation}
Here the coefficients $p_i$ and $q_i$ are to be determined from PT by  calculating $r(n)$ for $n=1,\ldots,2N$ and solving the resulting equations for $p_i$ and $q_i$. Note that Eq.~(\ref{eq:HAC}) is the defining property of hypergeometric functions $_{N+1}F_{N}$. 

\begin{figure*}
\includegraphics[width=0.95\textwidth]{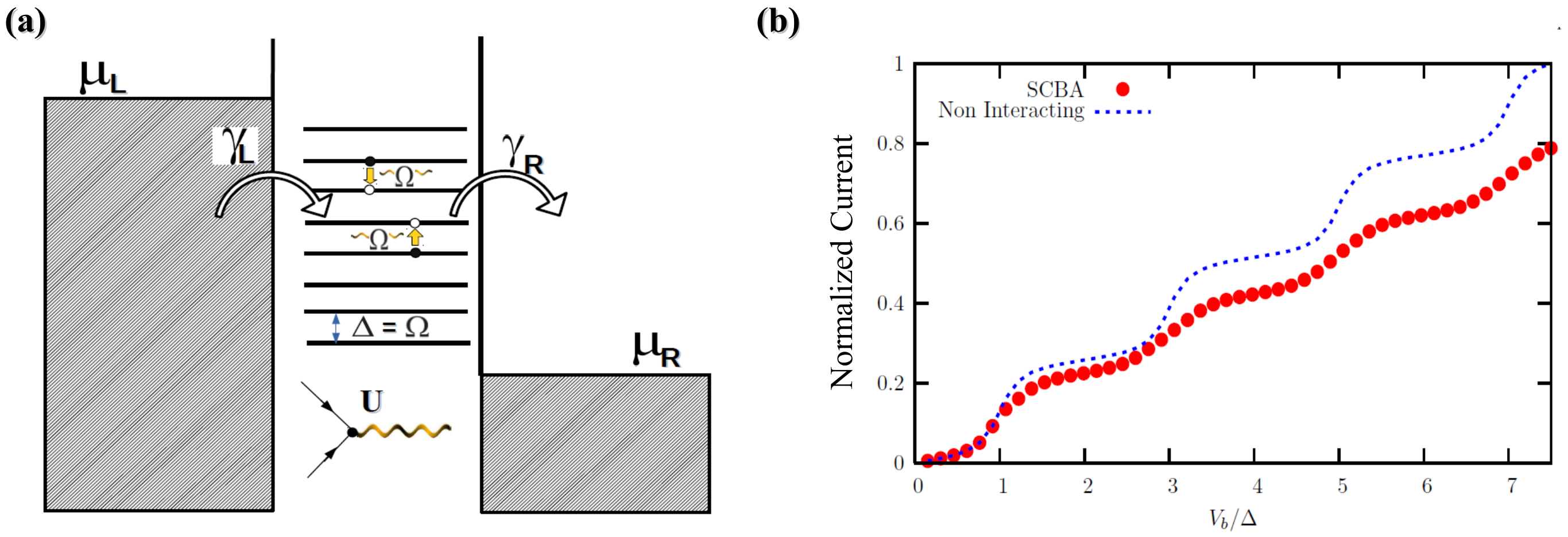}
\caption{(Color online) Schematic view of the single-molecule junction model driven by finite bias voltage $\mu_L-\mu_R = V_b$. Here $\gamma_{L/R}$ is the tunneling rate in and out of the molecular electronic energy levels; $\Delta$ is the energy gap between neighbouring levels; $\Omega$ is the phonon frequency of the single vibrational mode taken into account; and $U$ is the strength of electron-phonon interaction. (b) Current-voltage characteristics of the single-molecule junction in (a) in the absence ($U=0$, dashed line) and presence ($U=4.0 \gamma_{R}$, filled dots) of electron-phonon interaction. The latter case is computed using Fock-SCBA. The current is normalized by dividing by its non-interacting value at $V_b=7.5 \Delta$. For the model considered the interaction between electrons and phonon reduces the magnitude of the current relative to the non-interacting case.}
\label{fig1} 
\end{figure*}

The simplest rational approximation that generally tends to a non-zero constant as $n\rightarrow \infty$ is the 1/1 Pad\'e approximant. When applied to the ratio $\delta o_n/\delta o_{n-1}$ the 1/1 Pad\'e approximant yields the hypergeometric $_2F_1$. So, we approximate
\begin{equation}
r(n)=\frac{\delta o_n}{\delta o_{n-1}} \approx \frac{p_0+p_1 n}{1+q_1n}\,,
\label{eq:r11}
\end{equation}
where $p_0$, $p_1$ and $q_1$ are parameters left undetermined for the time being. This recursive equation can be solved to yield an analytic approximation for $o_n$, 
\begin{equation}
\delta o_n\approx \delta o_1 \frac{(-p_1/q_1)^{n-1}(2+p_0/p_1)_{n-1}}{(2+1/q_1)_{n-1}}\,,
\label{eq:on}
\end{equation}
in terms of the lowest order expansion coefficient $\delta o_1$ and the parameters $p_0$, $p_1$ and $q_1$. Now consider the perturbation expansion for the expectation value to infinite order, $\mathcal{O}_\infty=\sum_{i=0}^\infty \delta o_i U^{2i}$.  For $U<U_c$ one has $\mathcal{O}[G]=\mathcal{O}_\infty$ and,  therefore, one can replace $\delta o_i$ by the expression given in Eq.~(\ref{eq:on}), and then sum the resulting series to obtain $\mathcal{O}[G]\approx \mathcal{O}_0+_2F_1(1,2+p_0/p_1;2+1/q_1;p_1 U^2/q_1)\Delta \mathcal{O}_1$. Therefore, starting  from a power-series representation valid only for $U< U_c$, we have found an alternative representation valid also when $U>U_c$
\begin{equation}\label{eq:hyper2}
\mathcal{O}[G] \approx \mathcal{O}_0 + \, _2F_1(1,2+p_0/p_1;2+1/q_1;p_1U^2/q_1)\Delta \mathcal{O}_1.
\end{equation}
To find the values of $p_0$, $p_1$ and $q_1$, however, we need to calculate $\delta o_2/ \delta o_1$, $\delta o_3/ \delta o_2$, and $\delta o_4/\delta o_3$ numerically by means of Eq.~(\ref{eq:gnpt}). Alternatively one can solve the full Dyson equation and find $\mathcal{O}[G]$ as a function of $U$ for small $U$, and from this dependence one immediately extracts $\delta o_i$, $\delta o_1$, $\delta o_2$ and $\delta o_3$. In the latter case one can use available self-consistent solvers based on Eq.~\eqref{eq:iterative} to obtain the coefficients for the ACF, while in the former case one needs to implement Eq.~(\ref{eq:gnpt}). 

Once $\delta o_i$ are found from the perturbation series of the expectation value, one uses Eq.~(\ref{eq:r11}) to obtain 
\begin{eqnarray}
r(1)&=&\frac{\delta o_2}{\delta o_1}=\frac{p_0+p_1}{1+q_1},\\
r(2)&=&\frac{\delta o_3}{\delta o_2}=\frac{p_0+2p_1}{1+2q_1},\\
r(3)&=&\frac{\delta o_4}{\delta o_3}=\frac{p_0+3p_1}{1+3q_1},
\end{eqnarray}
which can be solved to find $p_0$, $p_1$ and $q_1$. Note that this procedure can be extended---by increasing the order of the Pad\'e approximant for the ratio---to generate a sequence of approximants involving higher-order hypergeometric functions. We emphasize that expectation values calculated using 
either of the two flavours of hypergeometric resummation will satisfy conservation laws they ought to satisfy on the proviso that a conserving approximation is chosen for $\Sigma$.

\section{Phonon-Limited Electronic Current in Molecular Junctions via Hypergeometric Resummation}\label{sec:junction}

In this Section we illustrate the potential of hypergeometric resummation as a convergence accelerator for the self-consistent solution of Eq.~\eqref{eq:dyson}. We consider current-voltage ({\it I--V}) characteristics of the model of single-molecule junction in the presence of electron-phonon interactions, treated at the level of Fock-SCBA. We use Eq.~\eqref{eq:gnpt} to evaluate the Fock-SCBA Feynman diagrams up to fourth order shown in Fig.~\ref{fig:fig0}, then compute expectation values from the resulting perturbative NEGF, and finally apply  hypergeometric resummation to find approximations to the fully self-consistent  expectation values. In the calculations shown below, this approach results in a speed-up of about one to two orders of magnitude relative to the  standard SCBA iteration based on Eq.~\eqref{eq:iterative}.

\subsection{Molecular Junction Model}
We wish to try hypergeometric resummation in a steady-state nonequilibrium regime for an electron-boson interacting model that is as simple as possible, while retaining sufficient physics for it to be nontrivial. For transparency of discussion, we focus on electrons interacting with phonons, but our analysis can be applied to electrons interacting with other types of boson quasiparticles such as magnons.~\cite{Mahfouzi2014}  A schematic view of such model, describing a single-molecule junction where electrons interact with a single phonon mode, is depicted in Fig.~\ref{fig1}(a). The model includes multiple phonon processes and has a built-in asymmetry, so that conservation laws are not guaranteed by symmetry.

We consider an isolated molecule  whose non-interacting Hamiltonian in the basis of molecular orbitals is diagonal, with eigenenergies $\epsilon_i$, $i=1,\dots,8$, as given by $\hat{h}_0=\sum_{i}\epsilon_i c_i^\dagger c_i$ where $c_i^\dagger$ ($c_i$) creates (annihilates) one electron in the molecular orbital $i$. The eigenenergies are assumed to be equally spaced, satisfying  $\epsilon_{i+1}-\epsilon_{i}=\Delta$. The unit of energy is set by $\Delta/2=1$, and we use $\hbar=1=e$ for simplicity. The noninteracting Hamiltonian is a diagonal matrix in the basis of molecular orbitals, which is taken to be $\hat{h}_0=\textrm{diag(-7,-5,-3,-1,1,3,5,7)}$. This Hamiltonian describes noninteracting central region within the NEGF approach.~\cite{Stefanucci2013,Haug2008} 

The central region is attached to semi-infinite ideal (i.e., with no disorder or many-body interactions) left (L) and right (R) electrodes which terminate into mascroscopic reservoirs at infinity where electrons are assumed to be thermalized at chemical potential $\mu_L$ or $\mu_R$, respectively. The L,R leads are accounted for by self-energies $\Sigma_{L,R}^r$. Here we chose a wide-band limit and assume that the coupling to the electrodes does not break the symmetry of the molecule, so that the L and R self-energies are diagonal matrices in the basis of molecular orbitals,  $(\Sigma^r_{L/R})_{i,j}=-i\gamma_{L/R}\delta_{i,j}/2$.  We choose $\gamma_L=0.08$ and $\gamma_L=0.1$. Since $\Delta/\gamma_R=20$, the broadening of the molecular orbital energies induced by tunneling in and out of the electrodes is much smaller than the separation between the energy levels. The current is driven through the junction by applying the electrochemical potential difference $\mu_L-\mu_R=V_b$. The bias voltage $V_b$ is applied symmetrically to the electrodes, i.e., $\mu_L=V_b/2$ and $\mu_R=-V_b/2$.

The non-interacting NEGF $g$ yields the usual lesser $g^<$, greater $g^>$, retarded $g^r$ and advanced $g^a$ GFs via the Lagreth rules.~\cite{Stefanucci2013,Haug2008} In steady-state transport regime, these GFs Fourier transformed to frequency are given by
\begin{eqnarray}
g^{r,a}(\omega)&=&[\omega I-h_0-\Sigma_C^{r,a}(\omega)]^{-1}, \\
g^{<,>}(\omega)&=& g^r(\omega)\Sigma_C^{<,>}(\omega)g^a(\omega).
\end{eqnarray}
where $I$ is the identity matrix and  $\Sigma_C^r(\omega)=\Sigma_L^r(\omega) + \Sigma_R^r(\omega)$.

From $g^{r,a}(\omega)$ and $g^{<,>}(\omega)$ we can calculate any one-electron expectation value. For example, the non-interacting electronic current flowing through the interface between lead $\alpha=L,R$ and the central region is given by
\begin{equation}\label{eq:current}
\mathcal{I}_\alpha[g]=\frac{1}{2\pi} \int d\omega \, \mathrm{Tr}\left[ \Sigma_\alpha^< (\omega) g^>(\omega)-\Sigma_\alpha^> (\omega)g^<(\omega)  \right],
\end{equation}  
where $\Sigma_{L,R}^<(\omega) = -2 i f_{L,R}(\omega) \mathrm{Im} \Sigma^r_{L,R}(\omega)$,  $\Sigma_{L,R}^>(\omega) = 2 i [1-f_{L,R}(\omega)] \mathrm{Im} \Sigma^r_{L,R}(\omega)$ and $f_{L,R}(\omega) = f(\omega-\mu_{L,R})$ is the Fermi function of macroscopic reservoirs. In the numerical calculations shown below we assume zero temperature $T=0$ which enters through the Fermi (or Bose-Einstein for phonons) distribution function.

\subsection{Iterating the Self-consistent Born Approximation}\label{sec:scba}

The electron-phonon interaction is assumed to be localized in the central region. We consider only one free phonon mode of frequency $\Omega=\Delta$, i.e., resonant with the gap between molecular levels. This means that an electron in level $i$ can 
make a transition to level $i \pm 1$ by emitting or absorbing a phonon. We assumme that the phonons do not interact with each other and or flow into the electrodes, so they are described by the Hamiltonian $\hat{h}_\mathrm{ph}=\Omega a^\dagger a$. Thus, electrons have no influence on the motion of phonons, but phonons are allowed to influence the electrons. The Hamiltonian describing electron-phonon interaction within the central region, which is added on $\hat{h}_0$, is given by
\begin{equation}
\hat{h}_\mathrm{int}=\sum_{i,j} M_{i,j} c^\dagger_{j}c_{i} (a^\dagger+a),
\end{equation}
where $a^\dagger$ ($a$) creates (annihilates) one phonon in the central region. The matrix elements of the interaction matrix $M$ are taken to be of the form $M_{i,j} = U\delta_{j,i\pm1}$, where $U$ is the interaction strength. Therefore phonon-induced electronic transitions are only between neighboring electronic levels. 

The electron-phonon interaction effects out of equilibrium are most often treated at the SCBA level.~\cite{Frederiksen2007,Lee2009a,Rhyner2014,Luisier2009} The SCBA can be further simplified to Fock-SCBA where non-crossing ``sunset''-diagram illustrated in Fig.~\ref{fig:fig0} are retained while  the Hartree diagrams are neglected.~\cite{Lee2009a} In applications one also often neglects the real part of the electron-phonon Fock diagram.~\cite{Valin2014} These approximations work under the assumption of small polaron shifts. Here we adopt these additional approximations to simplify considerably numerical calculations while capturing essential features of inelastic electron-phonon scattering on the electronic current. 

With these simplifications and approximations the lesser/greater and retarded self-energies that account for electron-phonon scattering within the electronic subsystem are given by
\begin{widetext}
\begin{eqnarray}
\Sigma_I^{<,>}(\omega) & = & M\left[ (n_B(\Omega)+1)G^{<,>}(\omega\pm \Omega)+n_B(\Omega)G^{<,>}(\omega \mp \Omega)\right] M, \\
\Sigma_I^{r}(\omega) & = & \frac{1}{2}\left[ \Sigma_I^{>}(\omega)-\Sigma_I^{<}(\omega)\right],
\end{eqnarray}
\end{widetext}
where $n_B(\Omega)$ is the Bose-Einstein distribution function. The retarded and lesser interacting NEGFs are given by
\begin{eqnarray}
G^{r}(\omega)&=&[\omega I-\hat{h}_0-\Sigma_C^{r}(\omega)-\Sigma_I^{r}(\omega)]^{-1},\\
G^{<}(\omega)&=& G^r(\omega)\left[ \Sigma_C^{<}(\omega)+\Sigma_I^{<}(\omega)\right] G^a(\omega).
\end{eqnarray}
The iterative solution of these equations is typically approached according to Eq.~\eqref{eq:iterative}---we start with $G\approx g$; evaluate $\Sigma_I^{<,>}$ and  $\Sigma_I^r(\omega)$; calculate new $G^{r}(\omega)$ and $G^{<}(\omega)$ which are in turn used to evaluate a new approximation to $\Sigma_I^{<,>}(\omega)$ and  $\Sigma_I^r(\omega)$. 

The current in the interacting case is obtained also from Eq.~\eqref{eq:current} by replacing non-interacting $g^{<,>}$ with $G^{<,>}$. Because the SCBA is a conserving approximation the steady-state current is a conserved quantity, $\mathcal{I}_L[G]=-\mathcal{I}_R[G]$. We use this fact as convergence criterion in the iteration of the SCBA equations, stopping the iteration when current conservation is violated by less than 0.001\% in two consecutive iterations. One can verify that  if $\gamma_L=\gamma_R$, current is automatically conserved as a result of the symmetry of the system. So to test current conservation it is necessary to choose $\gamma_L \neq \gamma_R$ in the calculations. We note that Eq.~\eqref{eq:iterative} leads to convergence issues already at low values of the interaction strength, independently of the convergence criterion. In order to speed-up the convergence we proceed as follows---at each value of $V_b$ we start from $U=0$ and increase it slowly until we reach a self-consistent solution for $G^{r}(\omega)$ and $G^{<}(\omega)$ at the selected finite $U$. 

The electronic current as a function of bias voltage $V_b$ is shown in Fig.~\ref{fig1}(b) for $U=4\gamma_R=0.4$. The inelastic electron-phonon scattering acts to reduce the current at large $V_b$ by about 20\% relative to its non-interacting value, while washing out the steps in the {\it I--V} characteristics of the non-interacting junction. 

\subsection{Perturbation Series for the Current}

Instead of the conventional approach outlined in Sec.~\ref{sec:scba}, one can use Eq.~(\ref{eq:gnpt}) to derive a conserving perturbation series for the current. This procedure requires $g_N^{<,>}$ which is obtained by applying Langreth rules
\begin{eqnarray}
g_N^< &=& g_{N-1}^<+\left(g\sum_{n=1}^N \Sigma_{n}\Delta g_{N-n}\right)^<\\
&=& g_{N-1}^<+g^r\sum_{n=1}^N \Sigma^<_{n}\Delta g^a_{N-n}\nonumber\\
&+&g^<\sum_{n=1}^N \Sigma^a_{n}\Delta g^a_{N-n}\nonumber\\
&+&g^r\sum_{n=1}^N \Sigma^r_{n}\Delta g^<_{N-n}.
\end{eqnarray}
This means that $g^<_N$ can be written as
\begin{equation}
g^<_N=g^<+\Delta g_1^<+\cdots+\Delta g_N^<,
\end{equation}
where $\Delta g_N$ is the $N$-th order correction to $g$ given by $g_N-g_{N-1}$, i.e., the sum of all the contributions containing $N$ phonon lines which is proportional to $ U^{2N}$. Given that the current is  a linear functional of $G$ we can write
\begin{eqnarray}
\mathcal{I}[g_N]&=&\mathcal{I}[g]+\mathcal{I}[\Delta g_1]+\cdots+\mathcal{I}[\Delta g_N]\\
&=&\mathcal{I}_0+\Delta \mathcal{I}_1+\cdots+\Delta \mathcal{I}_N\,,
\end{eqnarray}
which in the case studied here is the perturbation series for the current in SCBA.

\subsection{High-Bias Current Degradation}

The high-bias current degradation is an important figure of merit in nanoelectronic devices.~\cite{Rhyner2014} It gives a metric for the impact of the various scattering mechanisms on the magnitude of electronic current. Here we quantify the high-bias current degradation using
\begin{equation}
\textrm{Current Degradation}=100\left(1-\frac{\mathcal{I}}{\mathcal{I}_0} \right),
\label{eq:degradation}
\end{equation}
where both $\mathcal{I}$ and $\mathcal{I}_0$ are calculated in the high-bias regime that starts at $V_b/\Delta=7.5$. 

\subsubsection{SCBA vs. Perturbation Theory}

\begin{figure}
\includegraphics[width=0.45\textwidth]{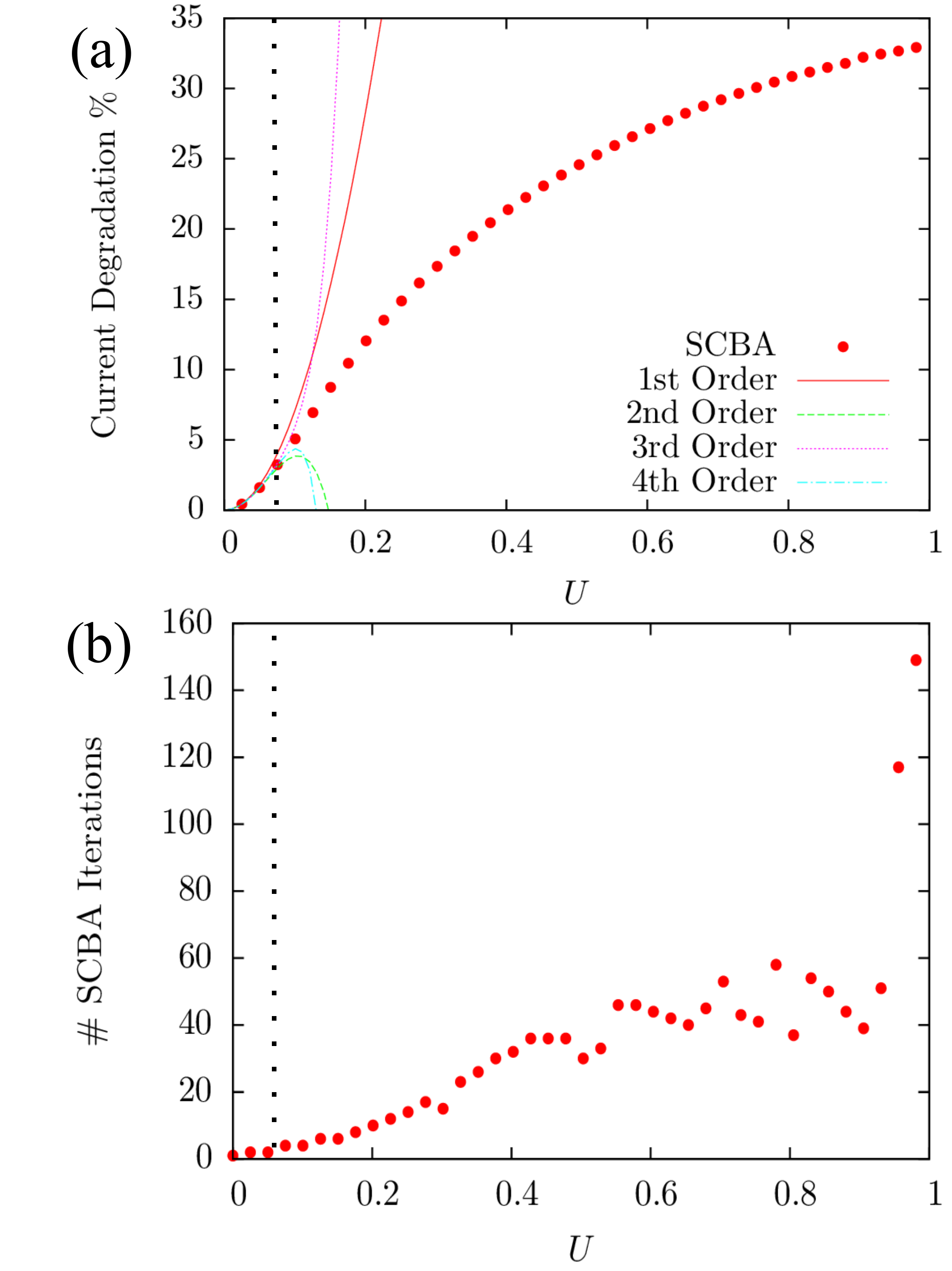}
\caption{(Color online) (a) Phonon-induced degradation of the electronic current as a function of the electron-phonon interaction strength $U$, calculated within the standard Fock-SCBA (dots) or perturbative Fock-SCBA (thin lines). Perturbation theory reproduces the self-consistent result only for very weak interactions as it diverges for $U>U_c\approx \gamma_L$. The perturbatively calculated current degradation is unphysical because it oscillates wildly as higher-order corrections are added to the series. In contrast, the standard Fock-SCBA approach based on Eq.~\eqref{eq:iterative} yields a regularized converged result. (b) Number of iterations of Eq.~\eqref{eq:iterative}  needed to converge the Fock-SCBA current. The vertical dotted black lines in both panels indicated the radius of convergence, $U_c \approx \gamma_L$, of the perturbation series for electronic current. The bias voltage is set as $V_b=7.5\Delta$.}
\label{fig:fig3}
\end{figure}

In Fig.~\ref{fig:fig3}(a) we show the calculated values of the current degradation as a function of $U$, for $0\le U \le1$. The value $U=1$ corresponds to $U/\gamma_R=10$ and, therefore, cannot be considered a weak interaction. The current degradation is calculated by means of the standard SCBA iteration based on Eq.~\eqref{eq:iterative}, as well as perturbatively by means of Eq.~\eqref{eq:gnpt}, using up to four orders of PT. The SCBA current degradation increases as a function of $U$, and reaches a value of about 35\% for $U=1$. We can see that perturbation theory does very poorly, oscillating wildly between odd and even orders. The perturbation expansion appears divergent outside of a small radius of convergence, $U_c \approx \gamma_L$ (note that $\gamma_L<\gamma_R$). Thus, the perturbative SCBA approach is therefore useless---the expansion is divergent and the self-consistent result is reproduced only for sufficiently small $U<U_c$. 

Figure~\ref{fig:fig3}(a) clearly illustrates the advantages of iterating the SCBA equations Eq.~\eqref{eq:iterative} over the perturbation expansion of Eq.~\eqref{eq:gnpt}---the standard iteration is in essence a resummation of the perturbation expansion for the NEGF that yields finite, well-behaved, {\it I--V} characteristics and current degradation in the presence of electron-phonon scattering. On the other hand Fig.~\ref{fig:fig3}(b) shows the potential problems with Eq.~\eqref{eq:iterative} which actually converges very slowly for $U>U_c$. As discussed in Sec.~\ref{sec:scba}, at each value of $U$ we input the self-consistent NEGF obtained from the previous value of $U$, which is slightly smaller, and yet some tens of  iterations are typically needed to reach self-consistency, e.g., to converge the NEGF for $U=0.2$ we start from the self-consistent solution obtained from $U=0.19$ and need  more than 10 iterations.

\subsubsection{Hypergeometric Resummation}\label{sec:hyperresum}

One can conclude that  Eq.~\eqref{eq:iterative} is not the most optimal choice to calculate expectation values from self-consistent MBPT. In Ref.~\onlinecite{Mera2013} it has been shown that a combination of MBPT with Pad\'e resummation can provide convergence acceleration relative to 
the standard SCBA iteration. In this Section, we demonstrate that for the example considered in Fig.~\ref{fig1} hypergeometric resummation 
provides near-ultimate convergence acceleration, outperforming both the standard SCBA approach and Pad\'e resummation.

\begin{figure}
\includegraphics[width=0.45\textwidth]{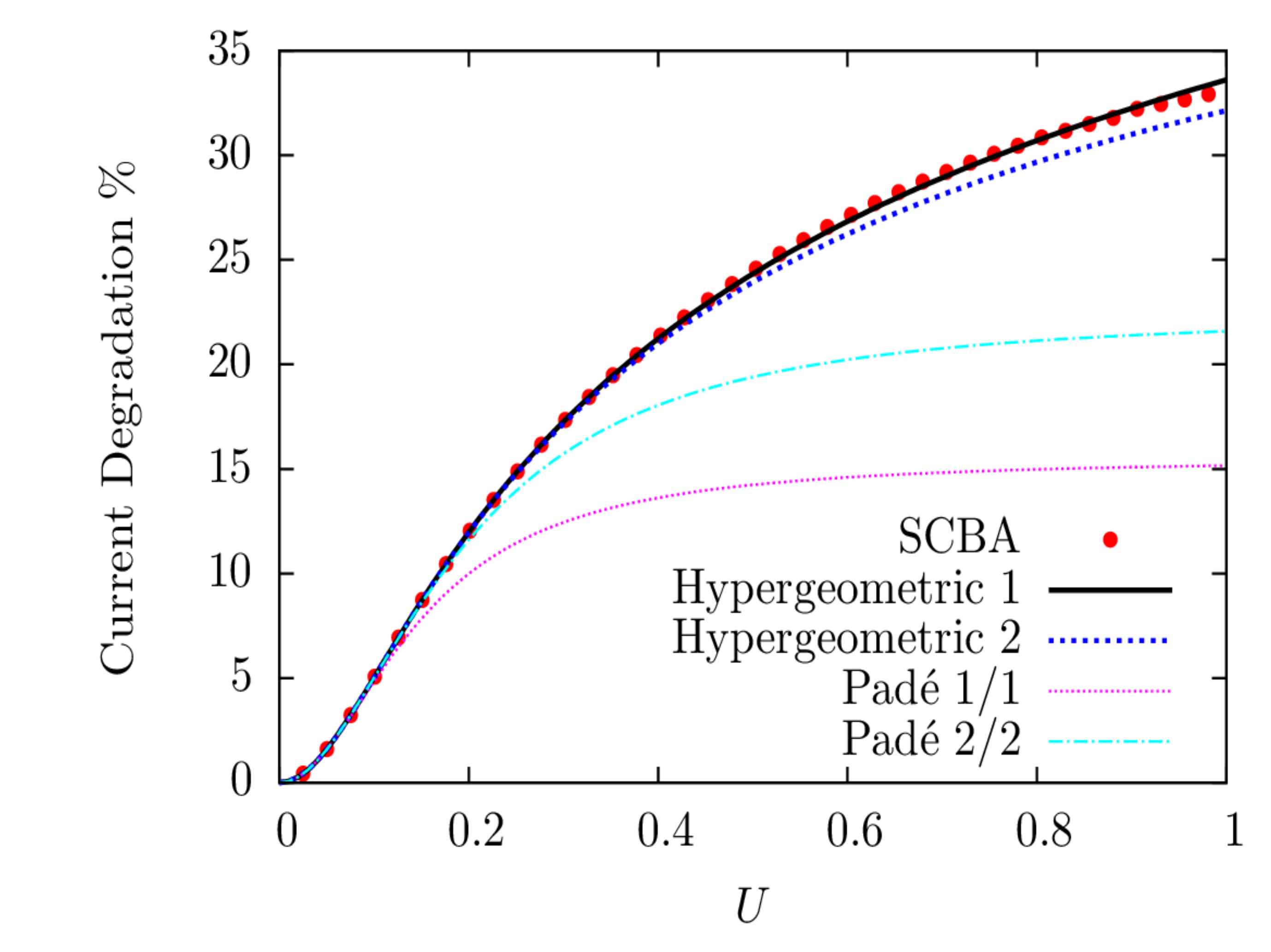}
\caption{(Color online) Phonon-induced current degradation as a function of $U$ calculated by the standard SCBA iteration given by: Eq.~\eqref{eq:iterative} (dots); the hypergeometric approximant given by Eq.~\eqref{eq:hyper1} (thick solid line); the hypergeometric approximant given by Eq.~\eqref{eq:hyper2} (thick dotted line); the 1/1 Pad\'e approximant (thin dotted line); and the 2/2 Pad\'e approximant (thin dot-dashed line). For the range of values of $U$ considered both hypergeometric approximants give excellent approximations to the self-consistent result, as well as a substantial improvement over Pad\'e approximants. The bias voltage is set as $V_b=7.5\Delta$.}
\label{fig:fig4}
\end{figure}

In Fig.~\ref{fig:fig4} we show the current degradation calculated as a function of $U$ using both flavours of hypergeometric resummation introduced in Sec.~\ref{sec:hyperapprox}, which are obtained from fourth order PT. Also shown are data obtained from the 1/1 and 2/2 Pad\'e approximants. Hypergeometric resummation essentially reproduces the fully self-consistent result, all the way up to $U=10\gamma_R=1$, and outperforms the second-order (1/1) and fourth order (2/2) diagonal Pad\'e approximants. It is quite remarkable how hypergeometric resummation manages to transform the fourth order PT result shown in Fig.~\ref{fig:fig3}(a) into the self-consistent result. The diagonal Pad\'e approximants also appear to work rather well, but they substantially underestimate the current degradation. These results, together with those of Ref.~\onlinecite{Mera2015}, suggest that hypergeometric resummation is potentially more powerful method than widely used Pad\'e approximants to sum a divergent PT using only a few terms.

The hypergeometric resummation provides excellent approximations to the fully self-consistent result, but at a minute fraction of the computational cost required~\cite{Frederiksen2007,Luisier2014} by standard SCBA. For instance, using Eq.~\eqref{eq:iterative} about thousands iterations are required to obtain the self-consistent result at $U=1$, while only four iterations of Eq.~\eqref{eq:gnpt} are required to build both hypergeometric approximants. Importantly, both hypergeometric and Pad\'e approximants  are exactly conserving, while in the standard SCBA conservation laws are never exactly obeyed due to the finite tolerances of practical calculations. It is worth  emphasizing that speed-up factors are highly dependent on the approach followed to iterate Eq.~\eqref{eq:iterative}. However, given their magnitude it appears unlikely that an approach based on Eq.~\eqref{eq:iterative} could do better than hypergeometric resummation for the examples considered here.

\subsubsection{I--V Characteristics by Hypergeometric Resummation at High Bias Voltage}

Finally, we turn our attention to the evaluation of the {\it I--V} characteristics in the presence of electron-phonon scattering for the single-molecule  junction model in Fig.~\ref{fig1}. In general the hypergeometric parameters of both approximants depend on the bias voltage $V_b$. Therefore one needs to perform a fourth order calculation for each value of $V_b$. However, in our calculations  we note that the hypergeometric parameters depend weakly on the applied $V_b$. This can be exploited to provide a very economical and accurate first-order approximation to the  electronic current in Fock-SCBA.  To do this, we determine the hypergeometric parameters at high $V_b$ and assume that all bias voltage dependence is contained in the non-interacting current or the first order current, depending on the hypergeometric approximant considered. 

\begin{figure}
\includegraphics[width=0.45\textwidth]{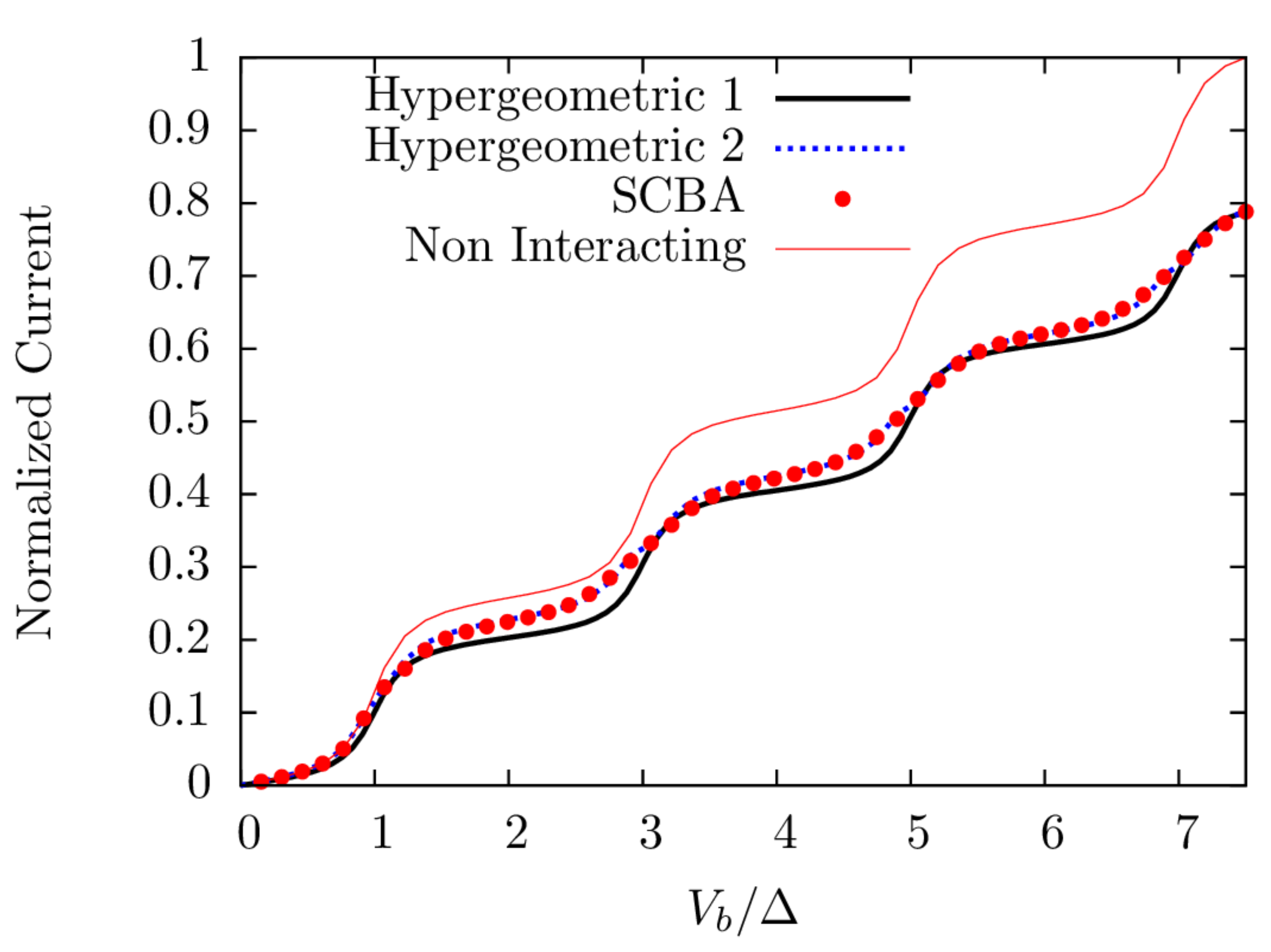}
\caption{(Color online) {\it I--V} characteristics calculated by hypergeometric resummation (thick solid and dashed lines) is compared with standard Fock-SCBA (dots). For reference, we give also the non-interacting current (thin dashed line). The current has been normalized to its non-interacting value at $V_b=7.5\Delta$. The value of the interaction strength is $U/\gamma_R=4$. The hypergeometric estimates were obtained in just 54 iterations of Eq.~\eqref{eq:gnpt}---a dramatic reduction relative to standard Eq.~\eqref{eq:iterative}.}
\label{fig:fig5}
\end{figure}

For this purpose, we use Eq.~\eqref{eq:hyper1} to obtain
\begin{equation}\label{eq:hypercurrent1}
\mathcal{I}(V_b) \approx \, _2F_1(h_1,h_2;h_3;h_4 U^2)\mathcal{I}_0(V_b),
\end{equation}
or we use Eq.~\eqref{eq:hyper2} to obtain
\begin{equation}\label{eq:hypercurrent2}
\mathcal{I}(V_b) \approx \mathcal{I}_0(V_b)+_2F_1(1,2+p_0/p_1;2+1/q_1;p_1U^2/q_1)\Delta \mathcal{I}_1(V_b).
\end{equation}
Here the coefficients ${h_i}$ and $p_0$, $p_1$, and $q_1$ {\it do not} depend on the bias voltage $V_b$. Applying these approximations to the model with parameter $U/\gamma_R=4$ yields the {\it I--V} curves shown in Fig.~\ref{fig:fig5}. The thin solid line gives the non-interacting {\it I--V} characteristics, with well defined steps in the typical staircase profile for transport through multilevel molecules. The red dots give the Fock-SCBA results obtained by iterating Eq.~\eqref{eq:iterative}. As the electron-phonon interaction strength, the current is degraded and the steps in the {\it I--V} staircase are washed out. The {\it I--V} curve eventually becomes linear function at sufficiently large $U$. Indeed, that  can be expected from a featureless (roughly constant) interacting electronic density of states (as obtained from the imaginary part of the interacting retarded GF). 

The thick solid line show results obtained from the hypergeometric approximant in Eq.~\eqref{eq:hypercurrent1}. We see that this hypergeometric approximant provides a very good estimate for the current, although the steps are sharper than their SCBA counterpart. That shows that the voltage dependence in the non-interacting current is not enough to describe the interaction-induced washing out of the step-like structure. In contrast, the hypergeometric approximant in Eq.~\eqref{eq:hypercurrent2} accounts better for this effect of electron-phonon interactions. The standard Fock-SCBA and ``hypergeometric 2'' {\it I--V} curves in Fig.~\ref{fig:fig5} are almost indistinguishable for moderate interaction strengths. Increasing $U>4\gamma_R$  (data not shown) leads to Fock-SCBA computed  {\it I--V} curves becoming nearly linear, while both hypergeometric approximants still exhibit some step-like structure and continue to give excellent estimates of the SCBA current. Note that there are 50 points in the {\it I--V} characteristics shown in Fig.~\ref{fig:fig5}, which require thousands of  iterations when using Eq.~\eqref{eq:iterative} and {\it only} 54 iterations when using Eq.~\eqref{eq:gnpt}.
 
\section{Discussion}\label{sec:discussion}

In this paper, we introduced an alternative approach to self-consistency and conservation laws for nonequilibrium electron-boson quantum-many body systems treated by the NEGF formalism.  Results of example calculations reveal that hypergeometric resummation~\cite{Mera2015} is a very promising approach to the summation of divergent series in MBPT, such as the Fock-SCBA. This technique is computationally much more efficient than standard iterators or resummation based on widely used  Pad\'e approximants. In this Section, we argue that perturbation expansion associated with Fock-SCBA  has a possibly very small radius of convergence, and that the singularity structure responsible for the divergence of the perturbation expansion is a branch cut. This observation can be  substantiated by means of Feynman diagram counting argument insired by Ref.~\onlinecite{Bender1976}.
\begin{figure*}
\includegraphics[width=1.\textwidth]{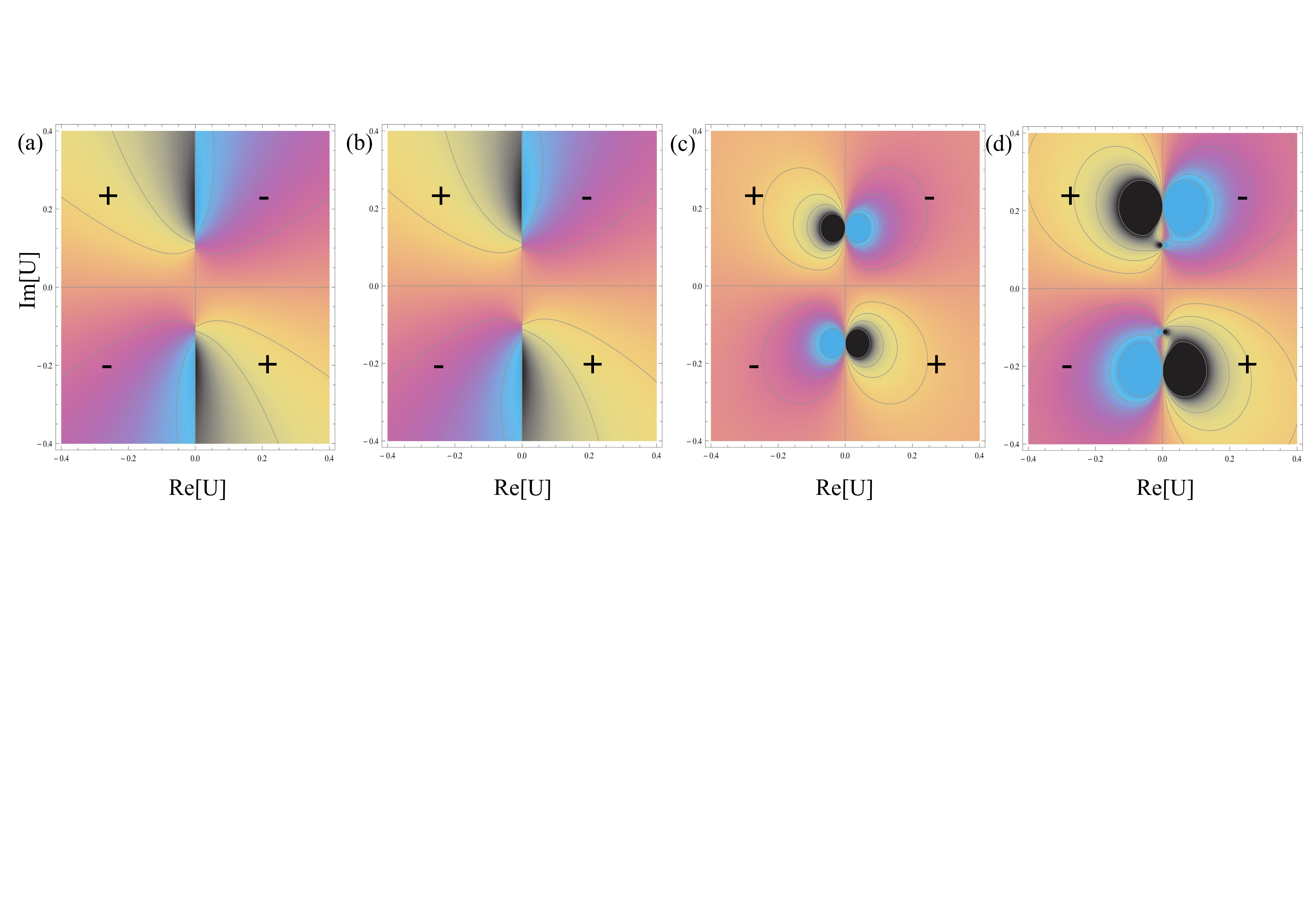}
\caption{(Color online) Imaginary part of the  phonon-limited electronic current at high bias voltage $V_b=7.5\Delta$ in the model of single-molecule junction from Fig.~\ref{fig1} as a function of complex $U$. The sign of the imaginary part of the current on each quadrant is indicated by $\pm$. The physical current is found for $\mathrm{Im} \, U=0$ where the imaginary part of the current is  zero. The current is calculated by summing the self-consistent ``sunset'' series in Fig.~\ref{fig:fig0} by means of: (a) the hypergeometric approximant given by  Eq.~\eqref{eq:hyper1}; (b) the hypergeometric approximant given by  Eq.~\eqref{eq:hyper2}; (c) the 1/1 Pad\'e approximant; (d) the 2/2 Pad\'e approximant. In (a) and (b) the convergence-limiting singularity is a branch cut along $\mathrm{Re}\, U=0$; the imaginary part of the current discontinuously changes sign across the imaginary axis. In (c) and (d) the convergence limiting singularity is the closest pole to the origin. Pad\'e approximants attempt to reproduce the cut by a line of poles---the 1/1 Pad\'e approximant has two poles, while the 2/2 Pad\'e approximant has four.}
\label{fig:fig6}
\end{figure*}

To understand hypergeometric resummation, it is useful to imagine that the electron-boson interaction strength $U$ is a complex parameter with both real and imaginary parts. In our example the physical system under consideration is recovered along the real axis. So let us consider the Fock-SCBA current as a function of complex $U$, $\mathcal{I}=\mathcal{I}(U)$. The observation of a divergent perturbation expansions in Fig.~\ref{fig:fig3}(a) signifies the presence of a singularity in the complex $U$ plane. The radius of convergence is given by the distance $U_c$ from the origin ($U=0$) to the nearest singularity in $\mathcal{I}(U)$.  The perturbation expansion for $\mathcal{I}(U)$ converges inside a circle $|U|<|U_c|$ and diverges in the annulus $|U| \ge |U_c|$, where  $U_c$ can be zero. The perturbation expansion is a polynomial that is unable to mimic the localized nature of a pole or a branch cut in the complex $U$-plane.

Pad\'e approximants are able to account for poles, but they are not well-suited to mimic branch cuts, unless one computes them to very large order. In Pad\'e resummation branch cuts are replaced by a string of poles. The higher the order of the Pad\'e approximant, the larger the number of poles used for the description of the branch cut and infinitely many poles are needed to precisely reproduce a branch cut. In contrast Gauss $_2F_1$hypergeometric functions have a built-in branch cut, as shown in Figs.~\ref{fig:fig6}(a) and ~\ref{fig:fig6}(b), and are also able to model poles. Thus, the fact that hypergeometric resummation outperforms Pad\'e approximants  suggests that the function  $\mathcal{I}(U)$ contains a branch  cut.

In Fig.~\ref{fig:fig6} we show imaginary part of the current  $\mathrm{Im}\,[\mathcal{I}(U)]$ in the complex $U$ plane, as given by both hypergeometric approximants and the 1/1 and 2/2 Pad\'e approximants. Clearly the hypergeometric approximants have a branc cut along the imaginary-$U$ axis, while the Pad\'{e} approximants have poles along the imaginary axis.  The imaginary part of the hypergeometric approximants changes its sign discontinuously across the imaginary-$U$ axis. In contrast, Pad\'e approximants do not have the ability to model branch cut,  but instead accumulate poles along the imaginary axis as more orders of perturbation theory are used. Therefore when the convergence-limiting singularity is a branch cut, Pad\'e approximants require many more orders of perturbation theory to be as accurate as fourth-order hypergeometric resummation.  


It should be noted that so far we have been able to infer the presence of the branch cut in $\mathcal{I}(U)$ \emph{a posteriori} through comparison with the fully self-consistent solution. One naturally wonders whether it would be possible to infer the presence of the branch cut \emph{a priori}. To see this we apply a counting argument, originally put forth in Ref.~\cite{Bender1976}, to the perturbation expansion of the self-consistent ``sunset'' diagramatic series shown in Fig.~\ref{fig:fig0} which lacks closed fermionic lines. Figure~\ref{fig:fig0} shows that there is one first-order diagram, two second-order diagrams, five third-order diagrams, fourteen at fourth order, etc. The sequence 1, 1, 2, 5, 14, 42, 132, \ldots, is known as  \emph{Catalan sequence}.  The $n$-th order self-consistent ``sunset'' series has $C_n$ terms, where $C_n$ is the $n$-th order Catalan number. Consider then the series
\begin{equation}
\sum_{n=0} C_n x^n,
\end{equation}
which is known to be asymptotic to the generating function 
$$
\frac{2}{1+\sqrt{1-4x}}=1+ \, _2F_1(1,3/2,3,4x)x,
$$
that contains a square-root branch cut and has the same form as the hypergeometric approximant in Eq.~\eqref{eq:hyper2} inspired by the ratio test of series convergence. 

Hypergeometric approximants are very flexible functions able to adapt to both cuts and poles. This should be particularly true when the notion of hypergeometric resummation is understood broadly. Indeed, the reader may ask: Why $\,_2F_1$ and not $(_2F_1)^{-1}$ or $\,_3F_2$? There are infinitely many hypergeometric approximants compatible with fourth order data. These constitute an approximant space that needs to be carefully explored. Extra information might be needed to select an optimal approximant out of this space.

\section{Conclusions}\label{sec:conclusion}

In conclusion, we have applied a very recently developed hypergeometric resummation~\cite{Mera2015,Pedersen2015,Sanders2015} to the calculation of physical observables in nonequilibrium electron-boson quantum many-body system.  In particular, we have considered the hypergeometric resummation of the non-crossing self-consistent ``sunset'' diagrammatic series (comprising Fock-SCBA) for the NEGF of electrons interacting with phonons in the presence of applied bias voltage which can drive this system far from equilibrium. We tested the approach by computing the {\it I--V} characteristics and phonon-induced degradation of the electronic current at high bias voltage applied to a model of single-molecule junction. Hypergeometric resummation of low orders of Fock-SCBA perturbation series  for the current reproduces the full self-consistent solution  at a fraction of the computational cost. The excellent performance of hypergeometric resummation strongly suggests the possibility that convergence of self-consistent ``sunset'' series is limited by branch cut singularity. The development of more general resummation approaches for equilibrium and nonequilibrium MBPT, inspired by hypergeometric resummation considered here, and their deployment for realistic modeling of experimentally relevant systems constitutes a very interesting challenge which we leave for future studies.

\begin{acknowledgments}
H. M. and B. K. N. were supported by NSF Grant No. ECCS 1509094.
\end{acknowledgments}


\begin{thebibliography}{10}

\bibitem{Kamenev2011}
A. Kamenev, {\em Field Theory of Non-Equilibrium Systems} (Cambridge University
  Press, Cambridge, 2011).

\bibitem{Haug2008}
H. Haug and A.-P. Jauho, {\em Quantum kinetics in transport and optics of
  semiconductors} (Springer-Verlag, Berlin, 2008).

\bibitem{Stefanucci2013}
G. Stefanucci and R. van Leeuwen, {\em Nonequilibrium Many-Body Theory of
  Quantum Systems: A Modern Introduction} (Cambridge University Press,
  Cambridge, 2013).

\bibitem{Brandbyge2002}
M. {Brandbyge} {\it et~al.}, Phys. Rev. B {\bf 65},  165401  (2002).

\bibitem{Taylor2001}
J. {Taylor}, H. {Guo}, and J. {Wang}, Phys. Rev. B {\bf 63},  245407
  (2001).

\bibitem{Palacios2002}
J.~J. {Palacios} {\it et~al.}, Phys. Rev. B {\bf 66},  035322  (2002).

\bibitem{Areshkin2010}
D.~A. Areshkin and B.~K. Nikoli\'{c}, Phys. Rev. B {\bf 81},  155450  (2010).

\bibitem{Pecchia2004}
A. {Pecchia} and A. {Di Carlo}, Reports of Progress in Physics {\bf 67},  1497
  (2004).

\bibitem{Lake1997}
R. Lake, G. Klimeck, R.~C. Bowen, and D. Jovanovic, Journal of Applied Physics
  {\bf 81},  7845  (1997).

\bibitem{Kubis2011}
T. Kubis and P. Vogl, Phys. Rev. B {\bf 83},  195304  (2011).

\bibitem{Rhyner2014}
R. Rhyner and M. Luisier, Phys. Rev. B {\bf 89},  235311  (2014).

\bibitem{Luisier2014}
M. Luisier, Chem. Soc. Rev. {\bf 43},  4357  (2014).

\bibitem{Aeberhard2012}
U. Aeberhard, Phys. Rev. B {\bf 86},  115317  (2012).

\bibitem{Locatelli2014}
N. Locatelli, V. Cros, and J. Grollier, Nat Mater {\bf 13},  11  (2014).

\bibitem{Manchon2009a}
A. Manchon and S. Zhang, Phys. Rev. B {\bf 79},  174401  (2009).

\bibitem{Levy2006a}
P.~M. Levy and A. Fert, Phys. Rev. B {\bf 74},  224446  (2006).

\bibitem{Mahfouzi2014}
F. Mahfouzi and B.~K. Nikoli\ifmmode~\acute{c}\else \'{c}\fi{}, Phys. Rev. B
  {\bf 90},  045115  (2014).

\bibitem{Viljas2005}
J.~K. Viljas, J.~C. Cuevas, F. Pauly, and M. H\"afner, Phys. Rev. B {\bf 72},
  245415  (2005).

\bibitem{Paulsson2005}
M. {Paulsson}, T. {Frederiksen}, and M. {Brandbyge}, Physical Review B {\bf
  72},  201101  (2005).

\bibitem{Frederiksen2007}
T. Frederiksen, M. Paulsson, M. Brandbyge, and A.-P. Jauho, Phys. Rev. B {\bf
  75},  205413  (2007).

\bibitem{Mera2012}
H. Mera {\it et~al.}, Phys. Rev. B {\bf 86},  161404  (2012).

\bibitem{Mera2013}
H. Mera, M. Lannoo, N. Cavassilas, and M. Bescond, Phys. Rev. B {\bf 88},
  075147  (2013).

\bibitem{Cavassilas2013}
N. Cavassilas, M. Bescond, H. Mera, and M. Lannoo, Applied Physics Letters {\bf
  102},    (2013).

\bibitem{Bescond2013}
M. Bescond {\it et~al.}, Journal of Applied Physics {\bf 114},  153712  (2013).

\bibitem{Thygesen2008}
K.~S. Thygesen and A. Rubio, Physical Review B (Condensed Matter and Materials
  Physics) {\bf 77},  115333  (2008).

\bibitem{Spataru2009}
C.~D. Spataru, M.~S. Hybertsen, S.~G. Louie, and A.~J. Millis, Phys. Rev. B
  {\bf 79},  155110  (2009).

\bibitem{Dash2010}
L.~K. Dash, H. Ness, and R.~W. Godby, J. Chem. Phys. {\bf 132},  104113
  (2010).

\bibitem{Dash2011}
L.~K. Dash, H. Ness, and R.~W. Godby, Phys. Rev. B {\bf 84},  085433  (2011).

\bibitem{Tandetzky2015}
F. Tandetzky, J.~K. Dewhurst, S. Sharma, and E.~K.~U. Gross, Phys. Rev. B {\bf
  92},  115125  (2015).

\bibitem{Lee2009a}
W. Lee, N. Jean, and S. Sanvito, Phys. Rev. B {\bf 79},  085120  (2009).

\bibitem{Mera2015}
H. Mera, T.~G. Pedersen, and B.~K. Nikoli\ifmmode~\acute{c}\else \'{c}\fi{},
  Phys. Rev. Lett. {\bf 115},  143001  (2015).

\bibitem{Pedersen2015}
T.~G. Pedersen, H. Mera, and B.~K. Nikoli\'{c}, arXiv:1510.01551  (2015).

\bibitem{Sanders2015}
S. Sanders, C. Heinisch, and M. Holthaus, EPL (Europhysics Letters) {\bf 111},
  20002  (2015).

\bibitem{Caliceti2007}
E. Caliceti {\it et~al.}, Physics Reports {\bf 446},  1  (2007).

\bibitem{Baker1996}
G.~A. Baker and P. Graves-Morris, {\em Pad\'{e} Approximants} (Cambridge
  University Press, Cambridge, 1996).

\bibitem{Nieminen1977}
R.~M. Nieminen, J. Phys. F: Metal Phys. {\bf 7},  375  (1977).

\bibitem{Kerker1981}
G.~P. Kerker, Phys. Rev. B {\bf 23},  3082  (1981).

\bibitem{Kresse1996}
G. Kresse and J. Furthm\"uller, Phys. Rev. B {\bf 54},  11169  (1996).

\bibitem{Baym2000}
G. Baym,  in {\em Progress in Nonequilibrium Green's Functions: Proceedings of
  the Conference “Kadanoff-Baym Equations: Progress and Perspectives for
  Many-Body Physics” Rostock, Germany, 20 – 24 September 1999}, edited by
  M. Bonitz (World Scientific, Singapore, 2000).

\bibitem{Baym1962}
G. Baym, Phys. Rev. {\bf 127},  1391  (1962).

\bibitem{Molinari2005}
L.~G. Molinari, Phys. Rev. B {\bf 71},  113102  (2005).

\bibitem{Molinari2006}
L.~G. Molinari and N. Manini, Eur. Phys. J. B {\bf 51},  331  (2006).

\bibitem{Gukelberger2015}
J. Gukelberger, L. Huang, and P. Werner, Phys. Rev. B {\bf 91},  235114
  (2015).

\bibitem{Weniger1989}
E.~J. Weniger, Computer Physics Reports {\bf 10},  189   (1989).

\bibitem{Weniger2010}
E.~J. Weniger, Applied Numerical Mathematics {\bf 60},  1429   (2010).

\bibitem{Dyson1952}
F.~J. Dyson, Phys. Rev. {\bf 85},  631  (1952).

\bibitem{Vainshtein2002}
A. Vainshtein,  in {\em Continuous Advances in QCD 2002: ArkadyFest}, edited by
  K. Olive, M. Shifman, and M. Voloshin (World Scientific, Singapore, 2002).

\bibitem{Bender1973}
C.~M. Bender and T.~T. Wu, Phys. Rev. D {\bf 7},  1620  (1973).

\bibitem{Suslov2005}
I.~M. Suslov, J. Exp. Theor. Phys. {\bf 100},  1188  (2005).

\bibitem{Luisier2009}
M. Luisier and G. Klimeck, Phys. Rev. B {\bf 80},  155430  (2009).

\bibitem{Valin2014}
R. Valin, M. Aldegunde, A. Martinez, and J.~R. Barker, J. Appl. 
  Phys. {\bf 116},  084507  (2014).

\bibitem{Bender1976}
C.~M. Bender and T.~T. Wu, Phys. Rev. Lett. {\bf 37},  117  (1976).

\end{thebibliography}

\end{document}